%% file: main.tex
\documentclass[a4paper,11pt]{article}
\usepackage{jheppub} 
\usepackage{lineno}

\input{include_settings}

\input{include_commands}

\title{A multi-event interface for next-to-leading order calculations in {MadGraph5\_aMC@NLO}}

\preprint{TIF-UNIMI-2026-3}

\author[a]{R. Frederix,}
\author[b]{S. Roiser,}
\author[c]{R. Sch{\"o}fbeck,}
\author[c,1]{Z. Wettersten\note{Corresponding author.},}
\author[d]{and M. Zaro.}

\affiliation[a]{Department of Physics, Lund University,\\
S{\"o}lvegatan 14A, 223 62 Lund, Sweden}
\affiliation[b]{European Organization for Nuclear Research,\\
Esplanade des Particules 1, 1217 Meyrin, Switzerland}
\affiliation[c]{Marietta Blau Institute for Particle Physics, Austrian Academy of Sciences\\
Wiesingerstra{\ss}e 4, 1010 Wien, Austria}
\affiliation[d]{Università degli Studi di Milano \& INFN Sezione di Milano,\\
Via Celoria 16, 20133 Milano, Italy}

\emailAdd{physics@wettersten.com}

\abstract{We detail the implementation of a multi-event interface for next-to-leading order (NLO) calculations in \madgraph{}, allowing tree-level scattering amplitudes for multiple phase space points to be evaluated in each call to the integrated NLO differential cross section during event generation. Additionally, a multithreaded implementation based on this multi-event interface where tree-level amplitudes are evaluated in parallel across multiple CPU threads is presented for the Monte Carlo generation of quantum chromodynamical (QCD) events. Although this work primarily concerns the implemented code, some algorithmic changes involving the order of the application of phase-space cuts and calls to different scattering amplitudes are included. The codebase currently supports multi-threaded execution, but these changes pave the way for continued data parallelism in the form of on-CPU SIMD instructions or SIMT GPU offloading. A study in the runtime fraction spent in different diagrammatic contributions across various processes suggests that NLO QCD event generation are computationally dominated by tree-level scattering amplitude evaluations, which we show are perfectly suited for data parallelisation.}

\begin{document}
\maketitle
\flushbottom

\section{Introduction}
\label{sec:intro}

While the statistical uncertainties associated with Monte Carlo (MC) event generation can be made arbitrarily small through an increase in the number of generated events, the precision of theoretical predictions in high energy physics are limited by the order at which the perturbative expansion used for evaluating hard scattering probabilities is taken. At the upcoming High-Luminosity Large Hadron Collider \cite{Schmidt_2016}, experimental measurements are expected to reach unprecedented levels of precision \cite{HSFPhysicsEventGeneratorWG:2020gxw}. It is thus pertinent to provide accelerated computations for event generation beyond the leading order (LO) contributions to provide relevant theoretical comparison data for the next generation of experimental measurements.

In recent years, there has been extensive work on speeding up unweighted hard scattering event generation, particularly through the use of data parallelism and hardware acceleration \cite{Kanzaki:2010ym,Hagiwara:2010oca,Hagiwara:2013oka,Bothmann:2021nch,Carrazza:2021gpx,Bothmann:2023gew,Cruz-Martinez:2025kwa,Valassi:2021ljk,Valassi:2022dkc,Valassi:2023yud,Hageboeck:2023blb,Valassi:2025xfn,Hagebock:2025jyk,valassi2025newgpudevelopmentsmadgraph}. However, these efforts have primarily targeted LO event generation. The next step from the phenomenological side is thus the application of these techniques and methods also to higher order contributions, starting with next-to-leading order (NLO) event generation.

Here, we present the first stage of this development for \madgraph{} (\mg{}) \cite{Alwall:2014hca}, a software suite for quantum field theoretical (QFT) phenomenological studies, which provides the generic generation of unweighted event samples at both LO and NLO. Recently, the \cudacpp{} plugin \cite{Valassi:2021ljk,Valassi:2022dkc,Valassi:2023yud,Hageboeck:2023blb,Valassi:2025xfn,Hagebock:2025jyk,valassi2025newgpudevelopmentsmadgraph} for \mg{} was released, which ports numerical LO scattering amplitude routines used for \mg{} event generation to data-parallel architectures such as \textit{single instruction, multiple data}- (SIMD-)enabled CPUs and \textit{single instruction, multiple threads} (SIMT)  GPUs. Preliminary investigations are being made into the possibility of extending \cudacpp{} also to NLO event generation. However, as the LO and NLO event generation codebases in \mg{} are largely disjoint, this and related work necessitates the implementation of an NLO multi-event interface, allowing for the evaluation of scattering amplitudes across an arbitrary number of phase space points within a single call to the computed differential cross section. Below, we present just such an interface.

To test the veracity of this NLO multi-event interface, a data-parallel proof of concept prototype was implemented using multithreading provided through OpenMP \cite{openmp} to perform the evaluation of tree-level scattering amplitudes in parallel. While such multithreading is unlikely to provide any performance gain in comparison to running several event generation executions in parallel, it does test the validity of the implementation and provides a baseline for the possible speed-up provided by future work into SIMD and SIMT parallelisation. Furthermore, as the underlying event generation algorithm itself has seen no changes, we see that the multi-event implementation perfectly reproduces the original sequential results when run without vectorisation up to bitwise precision, save for one small caveat.

This paper is structured as follows: in \cref{sec:amplitudes_at_nlo} we present the structure of the differential cross section $d\sigma$ at NLO, both from a qualitative description of the theoretical structure and a more pragmatic presentation of the numerical implementation in \mg{}. We then give a brief overview of the event generation algorithm in \cref{sec:evgen_algo}, detailing both the original implementation and the modified algorithm for our multi-event interface. In \cref{sec:runtime_comp} we give some minimal runtime profiles for the upstream NLO event generation module alongside runtime tests for the multithreaded implementation using this multi-event interface. The latter are additionally used to estimate the possible speed-up from the future implementation of data-parallel hardware acceleration; these measurements are used as a starting point for discussion about future development plans for the NLO event generation routines in \cref{sec:outlook}.

\section{Scattering amplitudes at higher orders}
\label{sec:amplitudes_at_nlo}

In this section we detail the complications that arise when evaluating hard scattering amplitudes at higher perturbative orders, both from a general analytic perspective in \cref{sec:sigmad_theory} and in the specific context of the \mg{} code in \cref{sec:sigmad_code}. These descriptions aim to give an idea of the issues faced when implementing data-parallel scattering amplitudes within \mg{} NLO event generation but are non-exhaustive; for more extensive details on the theoretical side, we point to standard QFT textbooks \cite{Ellis:1991qj,Peskin:1995ev,Zee:706825,Srednicki:2007qs} and in particular to documentation on local subtraction schemes \cite{Frixione:1995ms,Frixione:1997np,Catani:1996jh,Catani:1996vz,Catani:2002hc}, while the computational aspects for \mg{} are detailed in several papers \cite{Frederix:2009yq,Hirschi:2011pa,Frederix:2018nkq}.

Before going on, we shall standardise some terminology we use throughout this section and the rest of this paper. First, by the scattering amplitude $\m{}$ we refer to the complex hard scattering probability amplitude (also known as the ``matrix element''), i.e.\ the set of all relevant colour-ordered Feynman diagrams contributing to a given parton-level process at a given perturbative order. By the differential cross section $d\sigma$ we mean
\begin{align}
    d\sigma = f_1(x_1,\mu_F) \, f_2(x_2,\mu_F) \, \msq{} \, d\Omega,
\end{align}
with $f_i,x_i,\mu_F$ the parton distribution function (PDF) of initial-state parton $i$ evaluated at Bjorken fraction $x_i$ at factorisation scale $\mu_F$, and $d\Omega$ the generic phase-space measure (differential Lorentz-invariant phase space) of the phase space corresponding to a given process; i.e.\ the convolution of the absolute scattering amplitude squared and the PDFs across phase space. However, as PDF evaluation factorises from hard scattering and the phase-space measure can be chosen almost freely, we will for most discussion equivocate the differential cross section to the absolute scattering amplitude squared. As the differential cross section is discussed more frequently than the integrated total cross section, we will refer to the former as the cross section and the latter as the \textit{total} or \textit{integrated} cross section.

Additionally, we define the LO  cross section $\dsiglo{}$ to be the value of $d\sigma$ evaluated for the leading-order contribution for the scattering amplitude $\m{}$. The NLO cross section $\dsignlo$ is then given by the value of $d\sigma$ evaluated for scattering amplitude contributions \textit{up to and including} the NLO contribution, i.e.\ $\dsignlo{} = \dsiglo{} \, + \ldots $. In this context, we refer to the LO scattering amplitude as the \textit{Born contribution} $\mb{}$ to the cross section. For simplicity in terminology we disregard loop-induced processes, whose generic treatment is provided by \mg{} albeit disjointly from standard LO and NLO calculations (see section 2.4.2  \cite{Alwall:2014hca} as well as \cite{Hirschi:2015iia}).

\subsection{NLO scattering amplitudes in perturbative QFT}
\label{sec:sigmad_theory}

Integrated NLO cross sections can, in general, be written in the form
\begin{align}
    \int d\Omega \, \dsignlo{} = \int d \Omega^{(n)} \; \left(\dsiglo{} + \dsigv{}\right) + \int d\Omega^{(n+1)} \dsigr{},
\end{align}
where $\dsigv{}$ and $\dsigr{}$ refer to the virtual interference (loop) contributions and the real emission (real) contributions to the cross section, respectively, and $d\Omega^{(k)}$ denotes the $k$-dimensional phase space measure. Notably, NLO calculations necessitate integration also over ultraviolet and infrared regions of the phase space, and while the former can be treated term-by-term through renormalisation, by the KLN theorem the latter only cancel in the summation of the real and virtual contributions \cite{10.1143/PTP.17.401,Kinoshita:1962ur,PhysRev.133.B1549}. Removing these singularities is a complicated task, involving local subtraction schemes where a zero is added to the cross section in such a way as to make both the reals and virtuals independently locally finite. This can be written as
\begin{align}
     \dsigv{} + \dsigr{} \to \dsigv{}\, +\,  &\dsigctv{} + \dsigr{} - \dsigctr{},\\
     &\dsigctv{} = \int \domn{1} \dsigctr{},
\end{align}
where ``counterevent cross sections'' $d\sigma_{\mathrm{CT}}$ are defined such as to cancel in the integrated total cross section $\sigma_{\mathrm{NLO}}$ by setting the ``counter term'' (CT) for the virtual contributions to equal the CT for the real contributions integrated over the available phase space $\domn{1}$ for the real emitted parton\footnote{Note that these CTs have no relation to the MC CTs that arise when matching NLO hard scattering events to parton showers and arise from distinct sources. However, MC CTs are also directly related to the Born amplitudes, and acceleration of tree-level amplitudes thus also speeds up MC CT evaluations.}. The choice of this CT defines the local subtraction scheme, and
at NLO two different schemes are predominantly used: the Catani-Seymour dipole scheme \cite{Catani:1996jh,Catani:1996vz,Catani:2002hc}, which we forego discussing here; and the subtraction scheme by Frixione, Kunszt, and Signer (FKS) \cite{Frixione:1995ms,Frixione:1997np} used in \mg{}. Below, we provide a very brief, qualitative, and non-exhaustive description of FKS subtraction.

For simplicity, using the notation of \cite{Frederix:2009yq}, we assume a completely hadronic interaction such that singularities occur exclusively in the $\as{}$ expansion. For a given partonic real emission process, we label the number of \textit{massless} strongly interacting partons $n_L$ and the number of \textit{massive} strongly interacting partons $n_H$. We can then construct the set of all pairs of partons that may yield divergences in the $(n+1)$-body phase space as
\begin{align}
    \mathcal{P}_{FKS} = \left\{ (i,j) {\Large{|}}  3  \leq i \leq n_L + n_H + 2, 1 \leq j \leq n_L + n_H + 2, i\neq j\right\},
\end{align}
where $i$ labels a strongly interacting final-state parton and $j$ labels any strongly interacting external parton. The pairs $\mathcal{P}_{FKS}$ are called FKS pairs, and the integrated phase space can be partitioned into regions where only one of these pairs are problematic, i.e.\ regions where the real contribution diverges either in their soft limits $p_{i,j}\to0$ or in their collinear limit $\vec{p_i} \parallel \vec{p_j}$ (where by $p_k$ and $\vec{p}_k$ we mean the four- and three-momentum of external parton $k$, respectively). As such, CTs can be defined independently for each FKS pair by applying a partitioning function to the real contributions and summing over the resulting partitions, which we refer to as FKS sectors.

The details of the CTs are beyond the scope of this paper but are well-documented in the plethora of literature on the FKS scheme, and we shall limit ourselves to noting that they have $n$-body kinematics, i.e.\ they are evaluated akin to Born contributions. As such, NLO cross sections $\dsignlo{}$ can be written entirely in terms of Born, real, and loop Feynman diagrams, where each $(n+1)$-body ``real'' event is paired with an $n$-body counterevent.

\subsection{\madgraph{} helicity amplitudes at NLO}
\label{sec:sigmad_code}

In \mg{}, NLO scattering amplitudes are evaluated much akin to how we described them in \cref{sec:sigmad_theory}. The differential cross section routine \sigintf takes as input three arguments: an array of random numbers $x$ from which the phase space point is generated\footnote{Technically, three different phase space points are generated from the same array $x$, corresponding to the $n$-body contribution, the $(n+1)$-contribution, and the CTs (and potentially more, if several FKS sectors are explicitly summed over rather than sampled). For simplicity, we will refer to these as the same phase space point, as they are associated with the same call to \sigintf and are generated from the same $x$.}, the volume of the cell this point belongs to in the sampled distribution for $x$, and the current fold of the integration. We disregard the details of integral folding here, as it has no significant impact on the details of \sigintf itself.

The general program flow of \sigintf is as follows

\begin{enumerate}
    \item A random FKS sector is chosen for the current event.
    \item $x$ is mapped to a phase space point for the $n$-body contributions.
    \item If the Born momentum fails to pass $n$-body phase space cuts, jump to the $(n+1)$-body contributions in step 6.
    \item Couplings are evaluated and updated at the scale of the CT momentum.
    \item Born and loop contributions are evaluated at the Born momentum.
    \item $x$ is mapped to a phase space point for the real emission and the CTs.
    \item If CT momenta fail to meet $n$-body phase space cuts, jump to the real contribution in step 12.
    \item Couplings are evaluated and updated at the scale of the real emission momentum.
    \item MC CTs are evaluated at the Born momentum.
    \item Couplings are evaluated and updated at the scale of the CT momentum.
    \item Relevant FKS CTs are evaluated at the CT momentum.
    \item If real emission momentum fails to meet the $(n+1)$-body phase space cuts, jump to step 15.
    \item Couplings are evaluated and updated at the scale of the real emission momentum.
    \item Real emission contribution is evaluated at the real emission momentum.
    \item Contributions are returned for MC integration or unweighting, depending on whether we are performing the cross section integration or unweighted event generation.
\end{enumerate}

There are some additional intricacies omitted from the list above; e.g.\ several FKS sectors may be simultaneously integrated if they map to the same $n$-body momentum, and rather than evaluating the virtual interference term for each phase space point it is generally approximated using the Born contribution, with the full loop integral evaluated stochastically based on how well the approximation performs (minimum $0.5$\% of events). Besides such details, though, the above program flow describes how \sigintf is executed, with the contributions taking the forms of certain combinations of tree-level and loop Feynman diagrams.

Notably, this algorithm is highly branching and thus unsuited for lockstep data-parallel evaluation, i.e.\ the simultaneous execution of the same instruction across multiple data streams. Furthermore, in-software most of the data used in individual evaluations (such as momenta and couplings) is stored globally. For this reason, we have rewritten the \sigintf routine with two goals in mind: making it more modular, and making it more local. Particularly, noting that there is significant work in applying hardware acceleration to tree-level scattering amplitudes for the LO \cudacpp{} plugin, we factorised these from the rest of the calculations, as well as combining each new section of the algorithm into a (potential) loop over $n$ phase space points. This reworked algorithm looks as illustrated below, noting that a call to this ``vectorised'' version of \sigintf for $n$ phase space points corresponds to $n$ calls to the original version and perfectly reproduces the results of the original algorithm for $n=1$:
\begin{enumerate}
    \item A random FKS sector is chosen for the current $n$ events.
    \item For each $x$, all used momenta are generated and stored alongside running couplings.
    \item For each $x$, all tree-level amplitudes are evaluated.
    \item For each $x$, all evaluated amplitudes are accumulated through calls to the original evaluation routines which now take the corresponding amplitudes as input arguments.
\end{enumerate}

In the present implementation of the multi-event interface, phase space cuts are only applied after the amplitude evaluation loop, implying superfluous evaluations which end up being discarded. We elaborate on this topic in \cref{sec:evgen_algo,sec:runtime_comp} and discuss it with respect to continued development in \cref{sec:outlook}. However, we do note already here that by turning scattering amplitudes into arguments for the subroutines in the accumulation loop we automatically make perfect reuse of amplitudes which may enter several contributions. The upstream \mg{} codebase does have a simple cache system, but only stores one scattering amplitude at a time; consequently, if e.g.\ the Born amplitude is re-evaluated for a new momentum in one of the CTs, the original amplitude will be re-evaluated for the next CT. Thus, our multi-event interface makes perfect reuse of these $n$-body amplitudes without the implementation of a ``smarter'' caching system.

As a sidenote, the multi-event interface with $n=1$  reproduces all evaluated scattering amplitudes to bitwise precision when compared to the original codebase supplied with the same seed, with a singular caveat. To avoid writing to global data within the parallelisable evaluation loop, the identification of helicity-symmetric terms has been moved to a distinct \texttt{set\_goodhel} subroutine. The first few phase space points entering the evaluation loop will thus already have their identical helicity configurations noted and avoid superfluous calculations. The resulting amplitudes and weights hence go through different floating point operations, and vary up to floating point rounding precision from the corresponding phase space points when compared to the upstream code base. This leaves the first few phase space points differing from the reference in the final significant digit, which may result in slightly different phase space grids being generated. Consequently, the resulting event sample may differ slightly from the upstream codebase. We may want to set up more extensive validation tests in upcoming work, but for now we consider replicating the same event sample for the same phase space grids sufficient.

\section{Event generation algorithm}
\label{sec:evgen_algo}

As mentioned above, the main goal of this work is to provide a multi-event interface for NLO calculations, with the multithreaded implementation presented here just a proof of principle. As such, let us consider the structure of the full NLO unweighted event generation process as it is implemented in \mg{}. Below, the existing NLO integration and event generation routines in \mg{} are provided in 
\cref{sec:evgen_seq}, followed by a qualitative description of the modifications made to accommodate the multi-event interface in \cref{sec:evgen_vec}.

\subsection{Phase space integration and sampling}
\label{sec:evgen_seq}

Aside from Feynman diagram generation and corresponding HELAS-like routines, the codebase for NLO event generation in \mg{} is completely disparate from the one used for LO computations. NLO computations use a modified and extended version of the MINT program \cite{Nason:2007vt} --- originally developed for generic MC integration and unweighted sampling --- and the structure, despite significant developments, still maintains the fundamental structure of MINT.

At its core, \mg{} NLO event generation is explicitly split into three distinct steps: 1) grid generation, 2) cross section integration, and 3) unweighted event generation. The former two are largely equivalent in-software, with the main distinction being the output; the former splits the underlying distribution into uniform cells which are sampled through a uniform distribution to determine the ``volume'' of each cell with respect to the absolute value of the integrand, while the latter uses this weighted grid to perform the proper integral to define the distribution generated events are unweighted against. Event generation then runs roughly the same program flow as cross section integration, with the distinction that it returns whenever an event is ready to be unweighted. Algorithmically, 
\begin{algorithmic}
\Repeat
\State $d\sigma' \gets \texttt{signintF}$
\State $r \gets \text{rng}([0,1])$
\State $U \gets r\times \text{max}(d\sigma)$
\Until{$U < d\sigma'$}
\end{algorithmic}

In practice, these three steps are close to identical from the point of calls to \sigintf: In each case, a loop over $n$ random phase space points calls to \sigintf to evaluate the differential cross section for the generated phase space points. The differentiating factors are that in step 1 the phase space points are generated uniformly across the underlying distribution, while for step 3 the loop continues until an event has stochastically been chosen for unweighting. Furthermore, in the final step events are not added to the accumulated integral, but compared to the maximum of the integrand for the purpose of unweighting. As such, the considerations for a multi-event interface as detailed in \cref{sec:evgen_vec} are practically identical for the three modes, with some caveats for the unweighting procedure.

\subsection{Modified integrand calls for data-parallel amplitudes}
\label{sec:evgen_vec}

As previously stated, the procedures for grid generation, cross section integration, and event generation are almost identical. As such, we limit the description of the multi-event accommodations to that of cross section integration, plus a small comment on event unweighting and how it will need further extension for expansion to arbitrary numbers of events evaluated in parallel.

The original loop for cross section integration roughly reads as follows
\begin{algorithmic}
\State $I \gets 0$
\While{$n < \text{requested points}$}
\State $x \gets \texttt{get\_random\_x}$
\State $d\sigma \gets \texttt{sigintF}(x)$
\If {$\texttt{pass\_cuts}(d\sigma)$}
\State $n = n+1$
\EndIf
\State $I = I + d \sigma$
\EndWhile
\end{algorithmic}
with some additional stops to handle potential zero-valued integrands and integrals. Modifying this algorithm to account for a multi-event interface is simple: the random number generation simply needs to be looped over for $1\leq m \leq \texttt{vector\_size}$, and similarly the integral accumulation needs to be looped over each evaluated phase space point. In practise this is more convoluted, in that the generated random numbers and all the data used for accumulation needs an additional dimension appended with entries for each vector index $m$, but this constitutes relatively minor changes at the software level, especially when compared to the level of vectorisation necessary in \sigintf and its called subroutines.

One specific point of consideration is the final unweighting procedure during event generation, which will need further consideration for future development in hardware-accelerated event generation. Due to the assumption of sequentiality, the unweighting procedure also assumes that only one event is to be unweighted at a time. Again, this is not an algorithmic issue; rather than a boolean success check for returning from the event generation routine, an integer $1 \leq i$ can be returned denoting the number of events to unweight from the latest evaluated batch --- the complications arise in the surrounding infrastructure built on these assumptions, which will need extension for storing and writing multiple events at a time. Unlike the vectorisation applied in \sigintf whose bias is necessarily limited by \texttt{vector\_size} (which for on-CPU parallelism is unlikely to be problematic, due to small values of \texttt{vector\_size}), modifications to the unweighting procedure need to be considered with extreme care to ensure that unweighting efficiency does not limit any potential speed-up from the batched event generation calls.

\section{Runtime comparisons}
\label{sec:runtime_comp}

\begin{table}[t]
    \centering
    \begin{tabular}{c|c|c|c|c}
         Process & Borns & Linked Borns & Reals & Virtuals  \\ \hline
         $pp \to t \overline{t} $& 38.6\% & 4.1\% & 8.9\% & 1.9\%\\ 
         $pp \to t \overline{t} \,j$ & 49.0\% & 8.7\% & 18.6\% & 7.1\% \\  \hline 
         $pp \to \gamma\gamma$ & 35.4\% & 0.3\% & 5.2\% & 0.5\%  \\
         $pp \to \gamma\gamma \, j$  & 41.5\% & 0.5\% & 13.1\% & 2.5\% \\ \hline 
         $pp \to W^+ W^- $& 57.1\% & 0.3\% & 13.2\% & 1.1\% \\ 
         $pp \to W^+ W^- \, j$ & 55.4\% & 0.6\% & 27.0\% & 2.0\% \\ 
    \end{tabular}
    \caption{Runtime contributions for the different scattering amplitude routines for the full program chain to generate 10 000 unweighted events for $pp$ colliding to $t\overline{t}$, $\gamma\gamma$, and $W^+ W^-$ plus zero or one additional massless QCD jet, all evaluated at NLO QCD precision in \mg{} version 3.6.4. We have separated the runtime fraction spent in $n$-body tree-level scattering amplitudes from the evaluation of colour- and charge-linked Borns as the two are disjoint in the codebase, and the latter have not yet been moved to the multi-event interface in our implementation. Note that the runtime fractions here refer to the algorithmic routines called and not to the type of contribution they are used for --- the full evaluation of virtual contributions, for example, takes up a more significant fraction of the total runtime than this table shows, but comparatively little time is spent evaluating the loop integrals themselves.}
    \label{tab:nlo_runtime_fractions}
\end{table}

In this section, we provide runtime profiles for the upstream \mg{} NLO event generation module alongside a performance test of the multithreaded implementation based on our multi-event interface based on the number of simultaneously scheduled threads. While we do not expect multithreading to provide any performance benefit when compared to the upstream \mg{} NLO event generation implementation --- which already schedules jobs across all available CPU threads --- we can use the fraction of the runtime spent in tree-level amplitude evaluations alongside the performance of our multithreaded implementation to estimate the possible speed-up achieved through e.g.\ on-CPU SIMD parallelisation.

\Cref{tab:nlo_runtime_fractions} provides a runtime profile for NLO event generation in \mg{} version 3.6.4 as measured using perf \cite{perfwikiPerfLinux}. Specifically, it shows the fraction of total CPU runtime is spent in evaluation routines for the various types of contributions (Borns, linked Borns\footnote{The colour- and charge-linked Borns are evaluated using pre-calculated Feynman diagram amplitudes provided by the Borns, which explains their relatively small impact on runtime.}, reals, and virtuals) when generating 10 000 unweighted events at NLO QCD precision for the various processes listed. Note that these numbers do not represent what fraction of the runtime is spent on each type of contribution, as e.g.\ Borns also enter the virtual, CT, and possibly real contribution evaluations; for example, the full loop contribution takes up a far larger runtime fraction than shown here, but only as much time as shown is spent specifically on evaluating loop integrals.

In light of the fact that we already have much of the infrastructure to parallelise tree-level amplitudes, \cref{tab:nlo_runtime_fractions} is very promising. We see that for sufficiently complex processes the runtime is dominated by tree-level scattering amplitude evaluations. While these fractions are not as significant as those seen at LO (c.f.\ \cite{Hagebock:2025jyk}), they are significant enough to expect substantial speed-up to be provided by hardware acceleration.

\begin{figure}[t]
    \centering
    \includegraphics[width=0.49\linewidth]{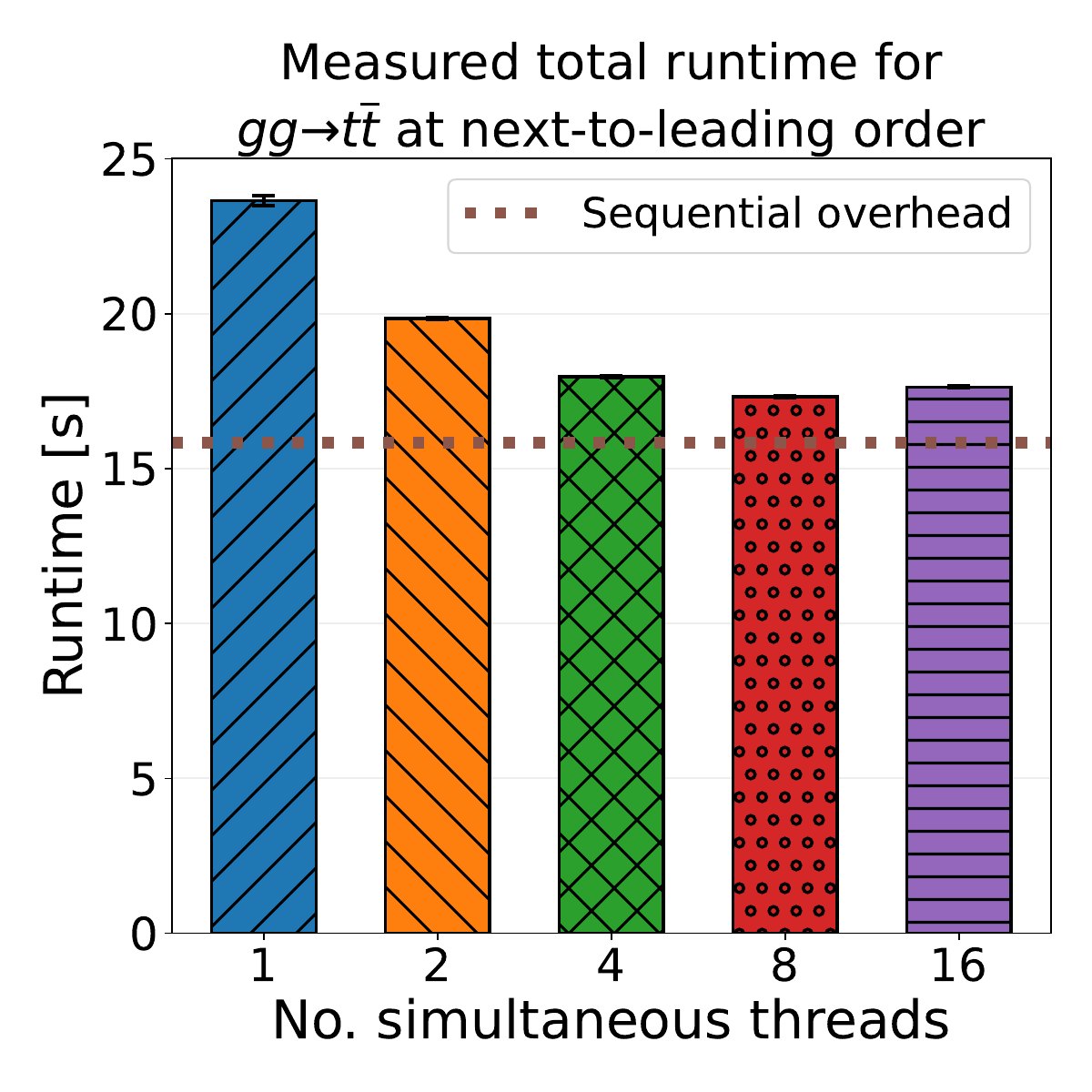}
    \hfill
    \includegraphics[width=0.49\linewidth]{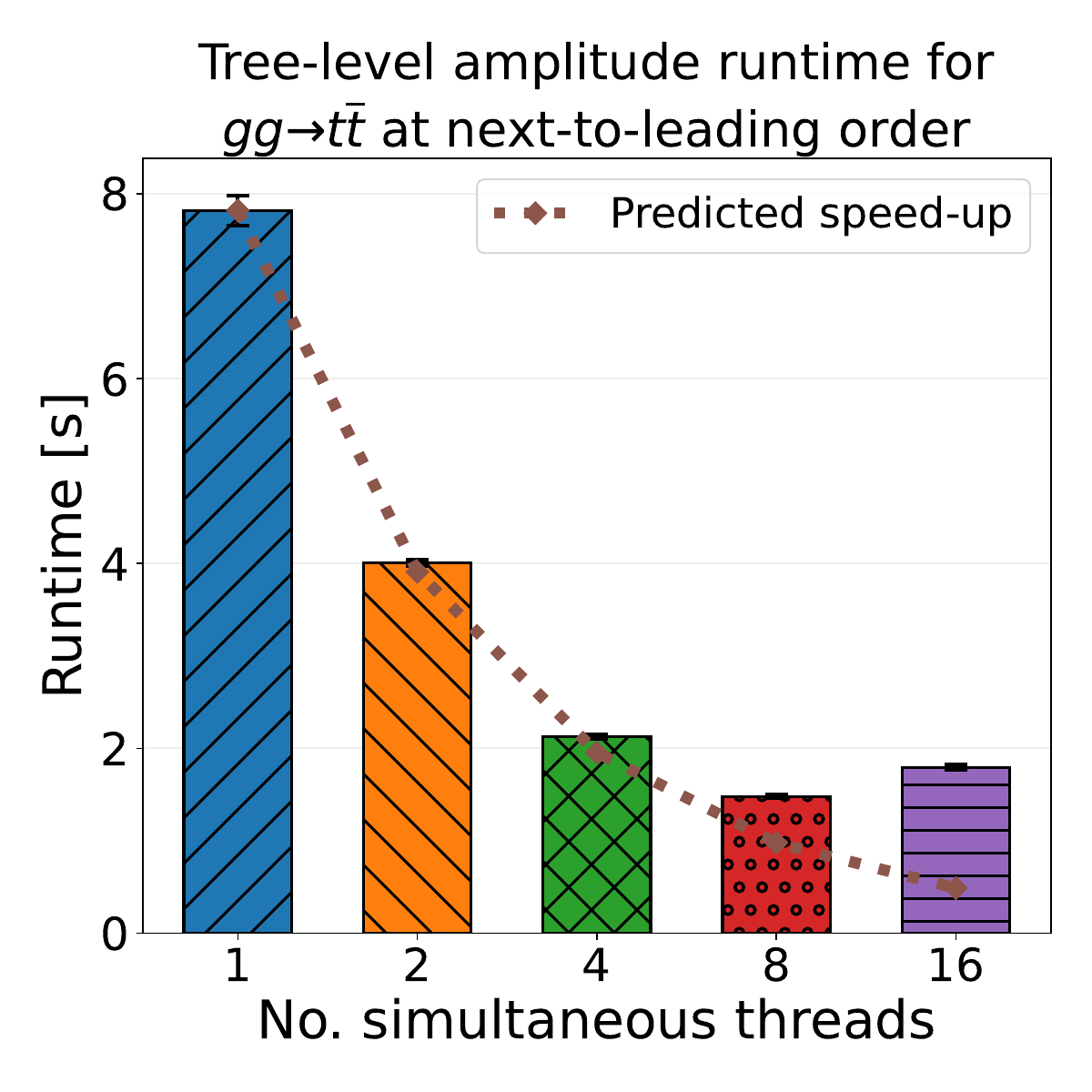}
    \caption{Runtime profile for one integration channel for the $g g \to t \overline{t}$ partonic contribution to the process $pp\to t \overline{t}$ in \mg{} --- using our multi-event interface --- as a function of the number of concurrent threads. For these tests, a single call was made to the differential cross section subroutine \texttt{sigintF} with 65 536 randomly generated phase space points to be evaluated through the multi-event interface, which greatly exaggerates the sequential overhead shown in the left-hand plot. The right-hand plot shows the real (wall) runtimes spent in scattering amplitude evaluations alongside the predicted runtimes, given by the single-threaded runtime divided by the number of concurrent threads. Displayed measurements are given by mean values of five independent runs, with standard deviations denoted by error bars. All tests were run on an AMD Ryzen 7 PRO 8840U.}
    \label{fig:omp_0j_runtimes}
\end{figure}

\begin{figure}[t]
    \centering
    \includegraphics[width=0.49\linewidth]{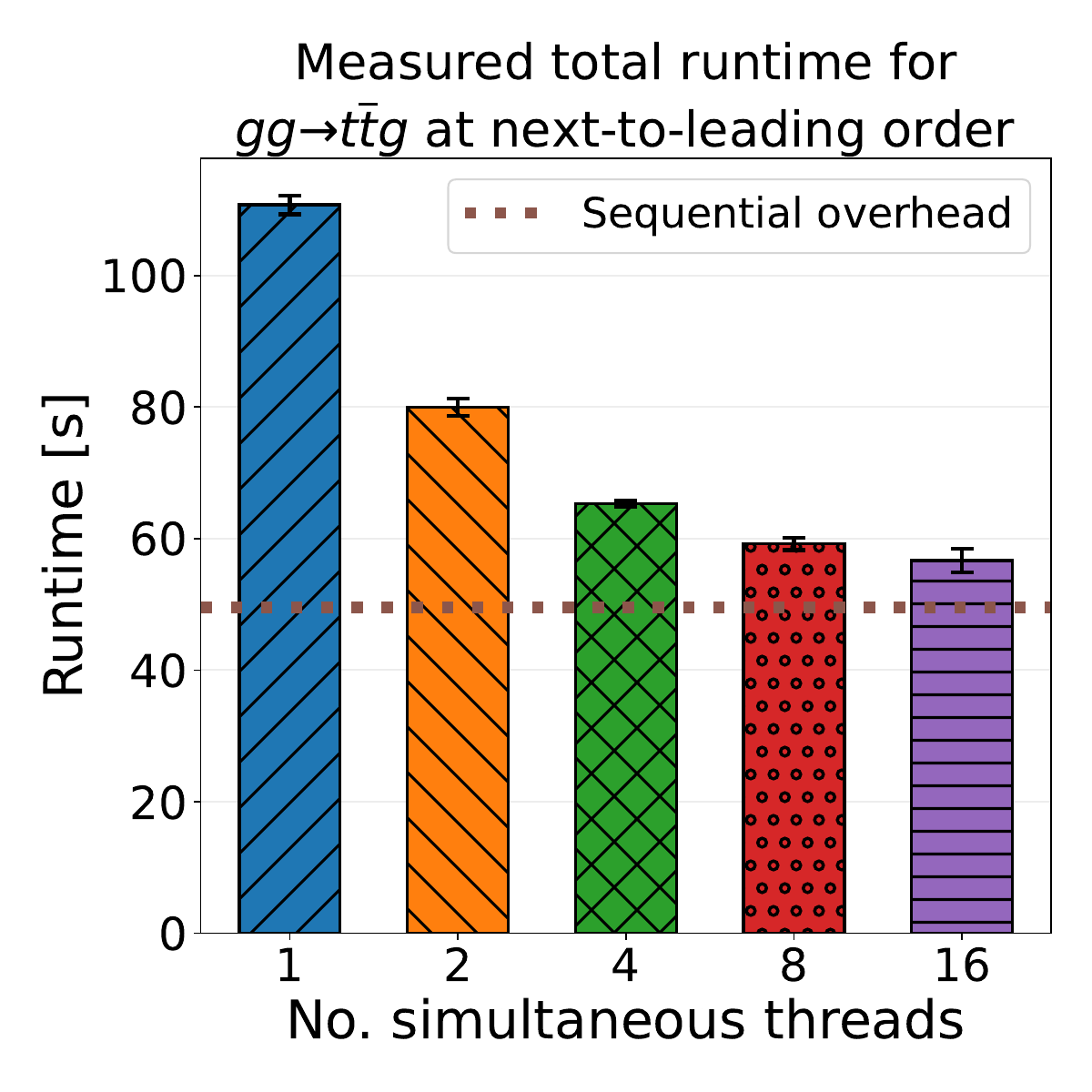}
    \hfill
    \includegraphics[width=0.49\linewidth]{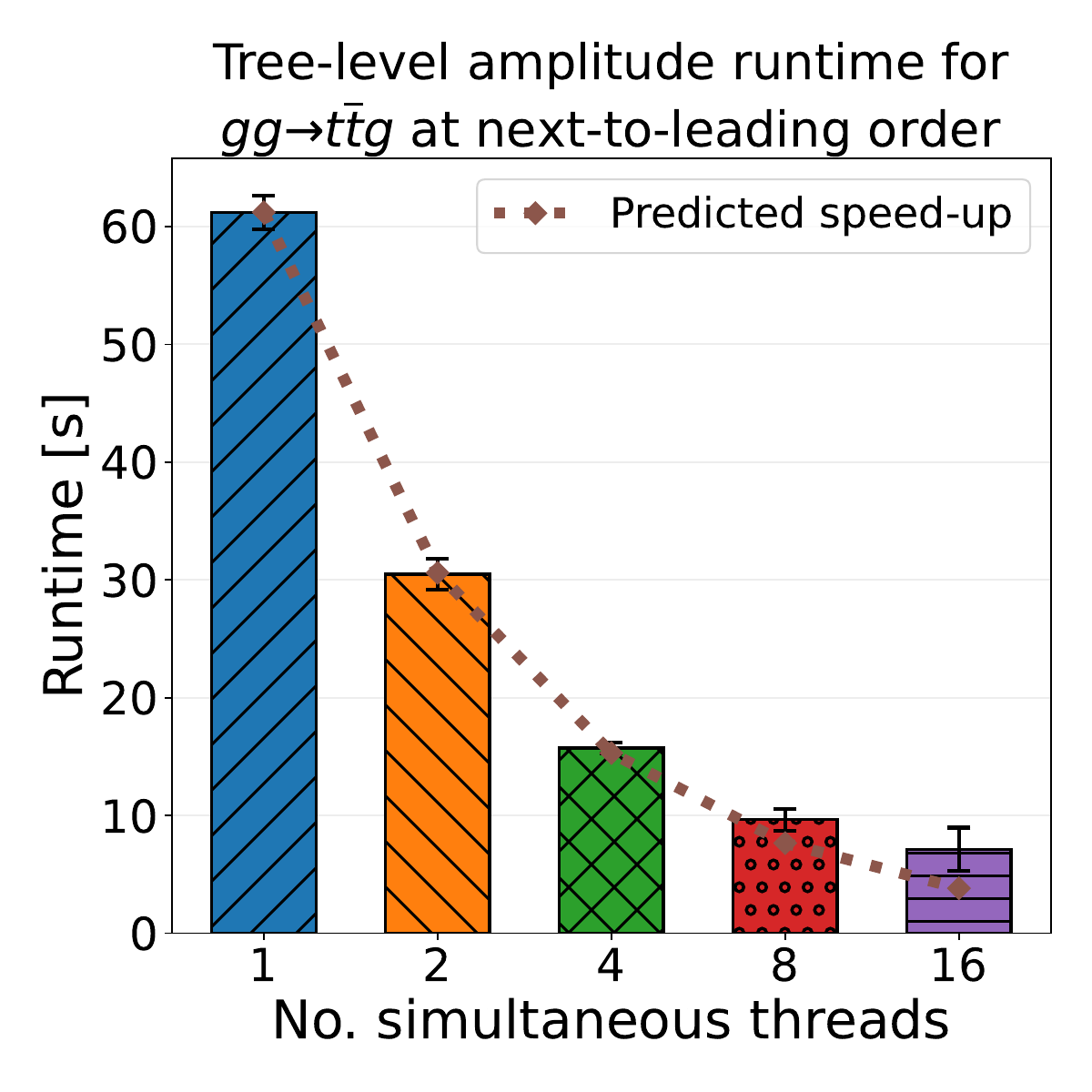}
    \caption{Runtime profile for one integration channel for the $g g \to t \overline{t}$ partonic contribution to the process $pp\to t \overline{t} \, j$ in \mg{} --- using our multi-event interface --- as a function of the number of concurrent threads. For these tests, a single call was made to the differential cross section subroutine \texttt{sigintF} with 65 536 randomly generated phase space points to be evaluated through the multi-event interface, which greatly exaggerates the sequential overhead shown in the left-hand plot. The right-hand plot shows the real (wall) runtimes spent in scattering amplitude evaluations alongside the predicted runtimes, given by the single-threaded runtime divided by the number of concurrent threads. Displayed measurements are given by mean values of five independent runs, with standard deviations denoted by error bars. All tests were run on an AMD Ryzen 7 PRO 8840U.}
    \label{fig:omp_1j_runtimes}
\end{figure}

Next, we turn to our multithreaded implementation. The present implementation only supports purely QCD processes, and as such we take the processes $p\,p \to t \overline{t} \,+ (0,1)j$ (with $p$ protons and $j$ massless QCD jets) as benchmarks. We first note that the overhead in the present multi-event interface extends total runtimes by roughly $25-35\%$, described primarily by the additional time spent in evaluating all possibly relevant amplitudes as well as overhead in setting global variables in the accumulation loop using existing sequential routines, leading to the same phase-space point-dependant variables being calculated multiple times for each point.

We now test the runtimes of multithreaded event generation with respect to the number of simultaneously scheduled threads. To simplify measurements, we limit these tests to the integration channel with the greatest contribution to the total cross section (i.e.\ the channel from which the most unweighted events are sampled), and due to the significant overhead in launching and accumulating threads we minimise the test to a single call to the differential cross section routine \texttt{sigintF} evaluating amplitudes for 65 536 random phase space points, which we leave the scheduler to launch evenly across CPU threads. This setup will exaggerate the computational overhead significantly in comparison to real applications, but ensures the benefit of data parallelism is apparent.

In \cref{fig:omp_0j_runtimes,fig:omp_1j_runtimes} the real (wall) runtimes for the full execution and exclusively the scattering amplitude routines is shown for one channel\footnote{By channels, we here refer to the different integrals in the single-diagram enhancement multi-channel integration method \cite{Maltoni:2002qb}, but this has little relevance for these tests.} of the gluonic parton configurations contributing to the two processes mentioned. Tests were run on an AMD Ryzen 7 PRO 8840U with eight cores and two threads per core, for a total of 16 threads. Shown numbers are mean runtimes across five runs for each displayed bar, with standard deviations in runtimes displayed as error bars. On the left-hand plots, the fraction of runtime spent outside the amplitude evaluation loop is denoted ``sequential overhead''; on the right-hand plots, ``predicted speed-up'' is given by the single-threaded evaluation loop runtimes divided by the number of concurrent threads.

The main takeaway from \cref{fig:omp_0j_runtimes,fig:omp_1j_runtimes} is clear: tree-level scattering amplitudes are perfectly suited for data parallelism also at NLO. Although speed-up does not increase perfectly with thread count in the many-threads regime --- even slowing down from 8 to 16 threads for the $g g \to t \overline{t}$ --- the fact that the more complicated $g g \to t \overline{t} \, g$ executable fits the predicted line far more closely suggests that this deviation is due to thread scheduling overhead rather than any issue with the implementation itself. Consequently, we can rather confidently assume that applying e.g.\ SIMD instructions rather than multithreading would provide linear speed-up in the evaluation loop with respect to the SIMD register size (i.e.\ a factor 4 (8) for AVX2 (AVX-512)). Again, we reiterate that the sequential overhead shown in \cref{fig:omp_0j_runtimes,fig:omp_1j_runtimes} is not representative of a real-world use case, but rather an extreme test used to showcase the validity of the multi-event interface.

With these points in mind, we can predict the expected possible speed-up from further developments to be at a minimum the runtime fractions of tree-level scattering amplitudes shown in \cref{tab:nlo_runtime_fractions} (noting that we believe that most of the introduced overhead in our multi-event interface can be removed in future optimisation). For e.g.\ $pp \to t \overline{t} \,j$, the total runtime could be reduced by $t_{\text{Borns}} \,+\, t_{\text{Reals}} \, = 67.6\%$; moving also colour- and charge-linked Borns into the multi-event interface would raise this number to $76.3\%$. We do not consider virtuals to be likely to be moved to this evaluation loop, at least not at present, and thus opt not to include them in this number. However, do note that there is ongoing work within multiple collaborations to also port loop integrals to data-parallel hardware, which may alter this choice within the near future.

\section{Outlook}
\label{sec:outlook}

We have presented a multi-event interface for NLO event generation in \madgraph{}, including both the factorisation and localisation of scattering amplitude calls within the integrated differential cross section and the treatment of surrounding phase space generation and integration. Although this work was limited to multithreaded data parallelism, the modified structure will relatively simply allow for future work following the form of the \cudacpp{} plugin, which enables data-parallel LO event generation in \mg{}, to implement on-CPU SIMD instructions and SIMT GPU offloading for tree-level contributions to the NLO cross section.

While on-CPU multithreading does not provide any performance benefits in comparison to running multiple executables in parallel, save for potentially speeding up single very time-consuming integration channels, the overhead introduced with our multi-event interface is minor, although not insignificant, at roughly $25-35\%$. We believe most of this overhead to be relatively easy to remove in future optimisation without any major restructuring in the multi-event interface necessary. Considering \cref{sec:runtime_comp} illustrates perfect data parallelism for tree-level scattering amplitudes, this suggests future work on hardware acceleration can minimise time spent on such evaluation and consequently speed up NLO event generation by roughly $50-75\%$ by replacing scattering amplitude routines with vectorised ones, e.g.\ those generated through the \cudacpp{} plugin.

There are some additional caveats to consider for future developments: the current implementation restricts all calls to the differential cross section in a single batch to belong to the same FKS sector. The low number of FKS sectors (generally around the single to low double digits) in comparison to a typical number of generated events (measured in thousands to millions of unweighted events, requiring orders of magnitude more generated weighted events depending on the unweighting efficiency) makes this a non-issue for the low level of parallel evaluations provided by on-CPU multithreading or SIMD instructions, but could significantly bias results when continuing onto e.g.\ GPU offloading where tens of thousands of events can be evaluated in parallel. Given the localisation that has already been performed in this work, extending the batching to consist of sub-batches which are expected to follow the same program path will be a minimal extension, but nevertheless one that needs to be implemented.

A more pressing issue is that the current implementation does not account for cuts within the vectorised cross section evaluation, i.e.\ \texttt{vector\_size} phase space points are sent to the evaluation routine at once, with cuts only applied at later stages. It is thus very possible that single threads corresponding to one of the phase space points may idle while the others run amplitude evaluations should it turn out that the point does not pass any phase space cuts. This will particularly impact lockstep processing for SIMD- and SIMT-acceleration and will need to be mitigated in some way. Handling this will involve both algorithmic and practical considerations: from the algorithmic side, complexity in the sorting algorithm when choosing what phase space points to evaluate in parallel must be decided with respect to the gain from parallelism (i.e.\ a trivial algorithm would loop over randomly generated numbers until $n$ phase space points all pass at least one cut, at which point \textit{all} contributions are evaluated) and memory considerations (e.g.\ a slightly but not significantly smarter algorithm which for each point checks which cuts the point passes and appends the point to a list of phase space points for which the corresponding amplitudes should be calculated, at the cost of storing multiple copies of the same phase space point). Practically, this will also necessitate the combination of random number generation and phase space mapping to allow for these to be done concurrently, unlike the current implementation where random number generation is performed before any call to the evaluation of the integrand where this random number is mapped onto the phase space and cuts can be processed. Similarly, the treatment of the stochastic evaluation of the full loop integral needs to be considered for lockstep processing. At present, the MC over full loop integral evaluation is left in the accumulation loop, but ongoing work in accelerating loop integrals makes it likely that we will want to move this, too, into the new evaluation loop.

Additionally, the unweighting routine may need more careful further consideration, particularly if we wish to continue with GPU offloading: the assumed sequential nature of integrand evaluations also extends to a practical assumption that at most one unweighted event will be generated per call to the differential cross section. While the development of a possible multi-event unweighting procedure is not inherently algorithmically problematic, it will necessitate the generalisation of a lot of code written assuming sequentiality. This will need to be done very carefully to not ruin either unweighting efficiency or physicality of resulting unweighted event samples.

Nevertheless, this multithreaded implementation and the multi-event interface developed for it is a first proper step towards the implementation of hardware acceleration for NLO event generation in \mg{}. For the time being, our results are bitwise identical to the original algorithm for \texttt{vector\_size}$=1$ while also providing statistically equivalent results for larger values of \texttt{vector\_size} so long as it is significantly smaller than the total number of generated events, illustrating that the algorithm is suited for event-level data parallelism. Future work may significantly accelerate NLO event generation through a now extant minimal multi-event interface between the original codebase and generic routines for scattering amplitude evaluations.

\acknowledgments
We extend our gratitude to Olivier Mattelaer for his assistance in understanding and modifying the codebase; to Stefano Frixione for discussions regarding the FKS subtraction scheme and its programmatic implementation in \madgraph{}; and to Andrea Valassi for help in sketching out the development plan and its more intricate details. We also thank all contributors to the \madgraph{} and \cudacpp{} projects, both past and present. Part of this work has been financed by the Next Generation Triggers project hosted by CERN, which is funded by the Eric and Wendy Schmidt Fund for Strategic Innovation. ZW ackowledges hospitality from the Physics Department of Milan University and INFN, Sezione di Milano, where part of this work has been carried out.




\printbibliography[heading=subbibintoc]

\end{document}

%% file: include_settings.tex
\usepackage[T1]{fontenc}

\usepackage{graphicx}%
\usepackage{multirow}%
\usepackage{amsmath,amssymb,amsfonts}%
\usepackage[title]{appendix}%
\usepackage{xcolor}%
\usepackage{textcomp}%
\usepackage{manyfoot}%

\usepackage[natbib,style=numeric-comp,sorting=none]{biblatex}
\usepackage[htt]{hyphenat}

\usepackage{booktabs}%
\usepackage{algorithm}%
\usepackage{algorithmicx}%
\usepackage{algpseudocode}%

\usepackage{soul}

\usepackage{multicol}
\usepackage{hyperref}
\usepackage{cleveref}
\usepackage{listings}

\usepackage{braket}

\usepackage{algpseudocode}

%% file: include_commands.tex
\newcommand{\cudacpp}{\texttt{CUDACPP}}

\newcommand{\madgraph}{\textsc{MadGraph5\_aMC@NLO}}
\newcommand{\mg}{\textsc{MG5aMC}}

\newcommand{\m}{\mathcal{M}}
\newcommand{\msq}{\left| \mathcal{M}\right|^2}

\newcommand{\dsiglo}{d\sigma_{\mathrm{LO}}}
\newcommand{\dsignlo}{d\sigma_{\mathrm{NLO}}}

\newcommand{\dsigv}{d\sigma_{\mathrm{V}}}
\newcommand{\dsigr}{d\sigma_{\mathrm{R}}}
\newcommand{\dsigctv}{d\sigma_{\mathrm{CT}}^{n}}
\newcommand{\dsigctr}{d\sigma_{\mathrm{CT}}^{n+1}}
\newcommand{\mb}{\mathcal{M}_{\mathrm{B}}}

\newcommand{\as}{\alpha_S}

\newcommand{\domn}[1]{d\Omega^{(#1)}}

\newcommand{\sigintf}{\texttt{sigintF}~}